\documentclass[aps,english,prb,reprint,superscriptaddress, 
citeautoscript,showpacs]{revtex4-1}
\usepackage{graphicx}
\usepackage{amssymb}
\usepackage{babel}
\usepackage{color}
\usepackage{multirow}
\usepackage{setspace}
\usepackage{longtable}
\begin{document}

\title{Two-dimensional pentagonal material \textit{Penta}-PdPSe: A first-principle study}

\author{A. Bafekry}\email{bafekry.asad@gmail.com}
\affiliation{Department of Radiation Application, Shahid Beheshti University, Tehran 1983969411, Iran}
\author{M. M. Fadlallah}
\affiliation{Department of Physics, Faculty of Science, Benha University, 13518 Benha, Egypt}
\author{M. Faraji}
\affiliation{Micro and Nanotechnology Graduate Program, TOBB University of Economics and Technology, Sogutozu Caddesi No 43 Sogutozu, 06560, Ankara, Turkey}
\author{A. Shafique}
\affiliation{Department of Physics, Lahore University of Management Sciences,Lahore, Pakistan}
\author{H. R. Jappor}
\affiliation{Department of Physics, College of Education for Pure Sciences, University of Babylon, Hilla, Iraq}
\author{I. Abdolhoseini Sarsari}
\affiliation{Department of Physics, Isfahan University of Technology, Isfahan, 84156-83111, Iran}
\author{Yee Sin Ang} 
\affiliation{Science, Mathematics and Technology (SMT) Cluster, Singapore University of Technology and Design, Singapore 487372}
\author{M. Ghergherehchi}
\affiliation{Department of Electrical and Computer Engineering, Sungkyunkwan University, 16419 Suwon, Korea}

\begin{abstract}

Low-symmetry Penta-PdPSe with intrinsic in-plane anisotropy synthesized successfully [P. Li et al., \textit{Adv. Mater.}, 2102541 (2021)]. Motivated by this experimental discovery, we investigate the structural, mechanical, electronic, optical and thermoelectric properties of PdPSe monolayer via density functional theory calculations. 
The phonon dispersion, molecular dynamics simulation, the cohesive energy mechanical properties of the \emph{penta}-PdPSe monolayer is verified to confirm its stability. 
The phonon spectrum represents a striking gap between the high-frequency and the low-frequency optical branches and an out-of-plane flexure mode with a quadratic dispersion in the long-wavelength limit. The Poisson's ratio indicates that \emph{penta}-PdPSe is a brittle monolayer.
The \emph{penda}-PdPSe monolayer is an indirect semiconductor with bandgap of 1.40 (2.07) eV using PBE (HSE06) functional. 
Optical properties simulation suggests that PdPSe is capable of absorbing a substantial range of visible to ultraviolet light. 
Band alignment analysis also reveals the compatibility of PdPSe for water splitting photocatalysis application. 
By combining the electrical and thermal transport properties of PdPSe, we show that a high $PF$ is achievable at room temperature, thus making PdPSe a candidate material for thermoelectric application. 
Our findings reveal the strong potential of \emph{penta}-PdPSe monolayer for a wide array of applications, including optoelectronic, water splitting and thermoelectric device applications.

\end{abstract}

\maketitle

\section{Introduction}
Two-dimensional (2D) materials have drawn much interests in recent years due to their superb optoelectronic, mechanical, and magnetic properties, extremely high surface areas, and narrow thickness at the nanoscale \cite{1,2,3,45}. It is well known that graphene is one of the most promising types of 2D materials thanks to its excellent Young's modulus and carrier mobility \cite{6,7,8}. However, the absence of a bandgap in graphene greatly limits its application in electronics device applications. As a material with configurable and suitable magnitude of band gap is much desirable for nanodevices applications, myriads of 2D materials have been fabricated and investigated one after the other in the past decade. Examples include graphene \cite{9}, phosphorene \cite{10}, silicene \cite{11}, antimonene \cite{12}, transition metal dichalcogenides (MoS${_2}$, WS${_2}$, WSe${_2}$, PtS${_2}$, and PtSe${_2}$) \cite{13,14}, metal dichalcogenides (SnSe${_2}$ and SnS${_2}$)\cite{15,16,17}, carbon nitrides (C${_3}$N, C${_3}$N${_4}$, C${_6}$N, C${_6}$N${_6}$ and C${_6}$N${_7}$) \cite{18,19,20}, Janus monolayers (Ga${_2}$SSe, Ga${_2}$STe, and Ga${_2}$SeTe) \cite{21}, monochalcogenides (GaSe, GaS, SnTe, InSe, and InS) \cite{22,23,24} and MA$_2$Z$_4$ (MoSi$_2$N$_4$ and WSi$_2$N$_4$) \cite{aa1, aa2, aa3, aa4}. These 2D materials have been widely explored for potential used in a colossal range of applications ranging from the storage and conversion of energy to electronics devices. 

Although the palladium phosphochalcogenides PdPX (X=Se, S) belonging to the category of layered inorganic materials and ternary pnictide chalcogenides have been characterized and synthesised their bulk materials since four decades ago \cite{25,26}, the 2D monolayer form of PdPX (X=Se, S) have received less attention in comparison with other 2D materials. Prior studies have revealed that the crystals structure of compounds PdPSe and PdPS are closer to that of PdSe${_2}$ and PdS${_2}$ \cite{27} and crystallized with orthorhombic structures and Pbcn space group \cite{28}. Besides, it was found that the PdPSe and PdPS compounds are semiconducting material with moderate band gaps of 1.28 eV \cite{29} and 1.45 eV \cite{30}, respectively. Recently, Jing et al. \cite{31} studied thin layers of PdPS and PdPSe with orthogonal lattices and demonstrated that the exfoliation of single-layer PdPX can be done from their bulk material with a cleavage energy of approximately 0.6 Jm$^{-2}$. They also confirmed that the thin layers of PdPS and PdPSe are kinetically, thermodynamically and mechanically stable. However, most of the described 2D materials have hexagonal or orthorhombic structures. 
Notwithstanding the tremendous potential of 2D materials in technological applications, obtaining stable 2D hexagonal structures remains a difficult task in practical devices. Consequently, there is an enormous growth of interests in producing 2D materials of different structures in order to cover a wider range of physical properties for various device applications. 
Particularly, 2D materials in \textit{pentagonal} lattice structures is relatively less explored compared to other high-symmetry lattices. 
Unlike the hexagonal systems, most \textit{penta}-2D materials such as \textit{penta}-SnS${_2}$ \cite{32}, \textit{penta}-graphene \cite{33}, and \textit{penta}-As${_2}$C \cite{34} are puckered and buckled in a regularly curved way to retain the symmetry. Amazingly, although most \textit{penta}-2D materials have been predicted theoretically \cite{35,36}, some of them have been recently tested experimentally. For example, 2D monolayer of PdSe${_2}$ with puckered pentagonal structure and isotropic demeanour have been characterized and fabricated by Oyedele et al \cite{37}. 

Very recently, 2D puckered \textit{penta}-PdPSe with a new low-symmetry pentagonal arrangement is experimentally exfoliated by Li et al \cite{38}. Surprisingly, Se and P atoms form an unusual [Se-P-P-Se]4- polyanions, which is currently the largest polyanion in the 2D systems. The robust intrinsic anisotropy of the \textit{penta}-PdPSe leads to unusual electronic and optical properties. The \textit{penta}-PdPSe layers were observed to have a high photoresponse of 5070 A/W at 635 nm, strong anisotropic electrical conductivity and an excellent electron mobility of 21.37 cm${_2}$V${^{-1}}$s${^{-1}}$. These results concretely establish \textit{penta}-PdPSe as a promising candidate material for building the next-generation of anisotropic electronics and optoelectronics device with novel functionalities. 
It should be noted that the wrinkled pentagons that make up the PdPSe-sublayers are uncommon in nature. 
Within the same sublayer, every Pd atom is linked to two Se atoms and two P atoms, producing a puckered tetracoordinate pattern. Hence, the atom coordinations of Pd, P and Se in \textit{penta}-PdPSe differ significantly from that of noble metal dichalcogenides and it is thus expected that \textit{penta}-PdPSe shall shed new lights on the physical properties of 2D materials not commonly found in 2D nobel metal dichalcogenides.

Despite the recent experimental breakthrough in the fabrication and the characterization of \textit{penta}-PdPSe, a comprehensive understanding on the mechanical, optical and electronic characteristics of \textit{penta}-PdPSe monolayer remains incomplete thus far. 
In this work, we attempt to fill in this gap by calculating the optical, electronic, transport and structural properties of the \textit{penta}-PdPSe monolayer using first-principle density functional theory (DFT) calculations. 
Based on the phonon dispersion, molecular dynamics simulation, and the cohesive energy mechanical properties, the stability of \emph{penta}-PdPSe monolayer is confirmed, and the calculated Poisson's ratio indicates that \emph{penta}-PdPSe is a brittle monolayer.
The \emph{penda}-PdPSe monolayer is an indirect semiconductor with bandgap of 1.40 (2.07) eV using PBE (HSE06) functional. 
PdPSe is predicted to absorb a substantial range of visible to ultraviolet light. 
Band alignment analysis also reveals the compatibility of PdPSe for water splitting photocatalysis application. 
We further show that a high $PF$ is achievable at room temperature, thus making PdPSe a candidate material for thermoelectric application. 
These findings shall pave the way for designing devices based on 2D \textit{penta}-PdPSe for a wide range of device applications. 

\begin{figure}[!b]
	\includegraphics[scale=1]{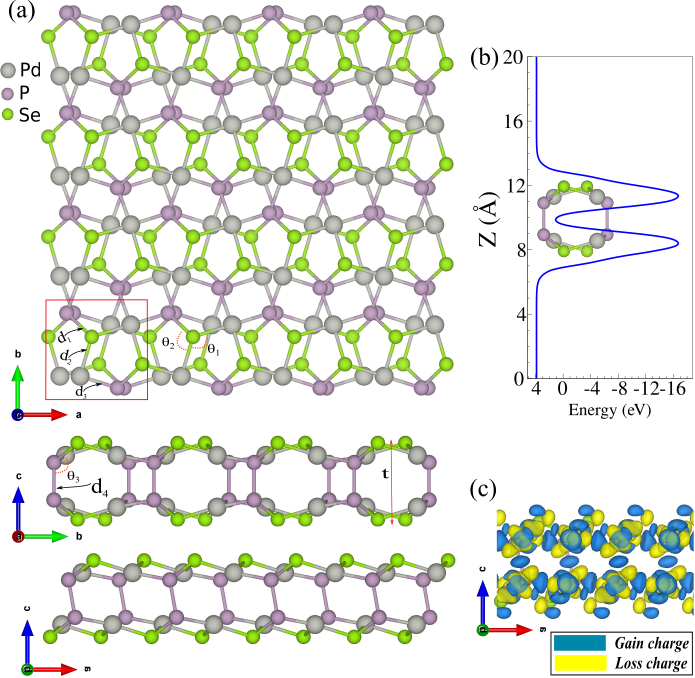}
	\caption{ (a) Atomic structure, (b) different charge density and (c) potential average of the PdPSe monolayer. 
		Rectangular primitive unit cell indicated by the red square. blue (yellow) 
		color refers to the accumulation (depletion) electron density.}
	\label{1}
\end{figure}

\section{Method}
The first-principles electronic structure calculations is performed within the density functional theory (DFT) defined in the Vienna \textit{ab-initio} simulation package (VASP) \cite{vasp1,vasp2}. To conduct the DFT calculations, the plane-wave basis projector augmented wave (PAW) method accompanied with the Perdew-Burke-Ernzerhof(PBE) \cite{GGA-PBE1,GGA-PBE2} is imposed. Furthermore, the Heyd-Scuseria-Ernzerhof (HSE06)\cite{hse} functional is employed for more accurate calculation of the bandgap value. 
We adjust the kinetic energy cut-off equal to 500 eV for the plane-wave expansion, then the energy of the system is converged till fluctuations in the energy is below 10$^{-5}$ eV. Optimization of the structures is performed with the total Hellmann-Feynman forces are decreased to 0.05 eV/\AA {}. 
21$\times$21$\times$1 $\Gamma$ centered \textit{k}-point grid was employed for the primary unit cells via taking the Monkhorst-Pack \cite{Monkhorst}. 
Charge migration analysis is performed by utilizing the Bader method \cite{Henkelman}. 
Here, the long-range vdW interactions can be explained by the Van der Waals (vdW) corrections \cite{Grimme}. 
The phonon characteristics were derived from the small displacement technique as employed in the PHONOPY code \cite{phon}.
The thermal stability is also examined by conducting the AIMD simulations at 600K.The training set is prepared by conducting ab-initio molecular dynamics (AIMD) simulations over $3\times3\times1$ supercells with $2\times2\times1$ k-point grids. AIMD simulations are carried out at 600K, each for 5000 time steps. 
The electronic transport coefficients are calculated using the Boltzmann transport theory under the rigid band and constant
relaxation approximation. 

The transport transport coefficients are calculated using the Boltztrap2
code \cite{boltztrap2}. The Monkhorst--Pack k-mesh of $30\times30\times1$
is used to accurately calculate the electronic transport coefficients.

\section{Results and Discussions}
\subsection{Structural and Mechanical properties}

\begin{figure}[!b]
\centering
\includegraphics[scale=1]{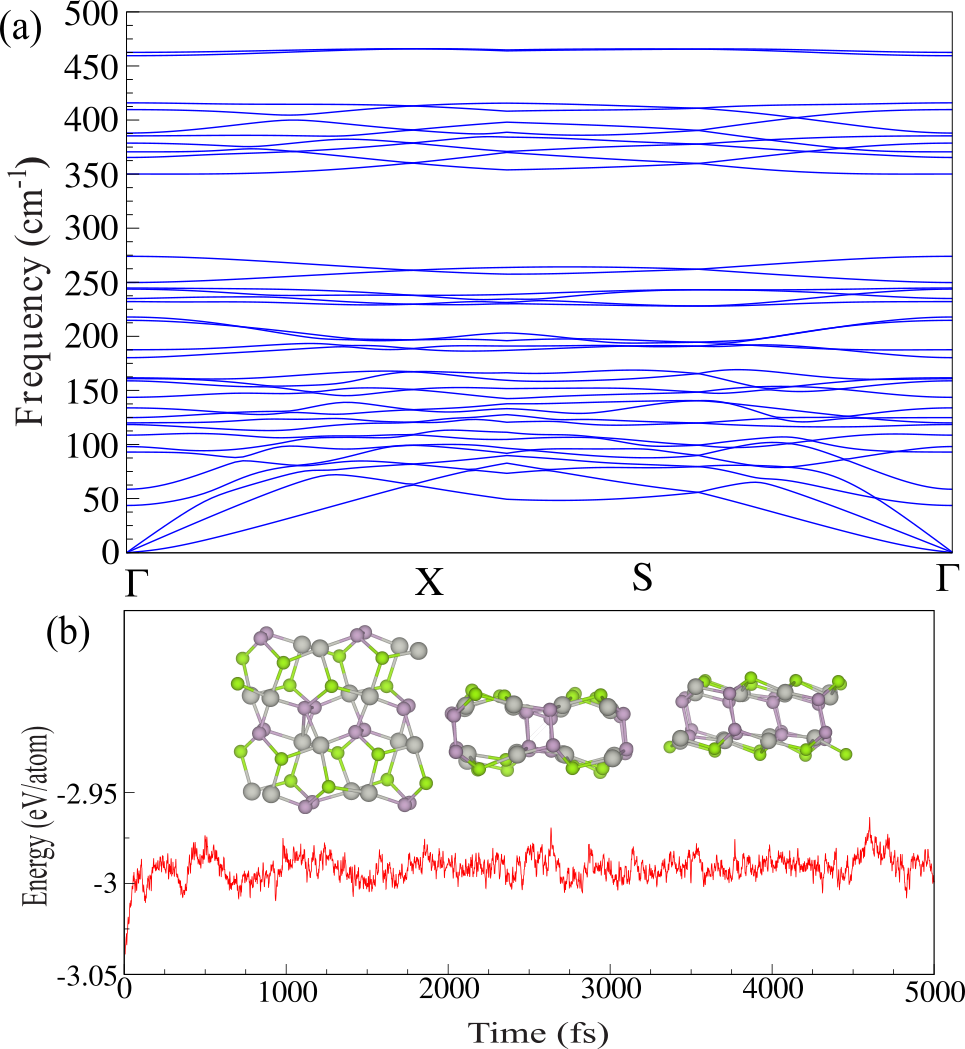}
\caption{ (a) Phonon band spectrum of PdPSe monolayer. (b) Ab initio molecular dynamics (AIMD) for PdPSe monolayer at 600 K temperature. Top view of the structure after 5 ps of simulation demonstrated as insets.}
\label{2}
\end{figure}

\begin{figure}[!htb]
	\centering
	\includegraphics[scale=1]{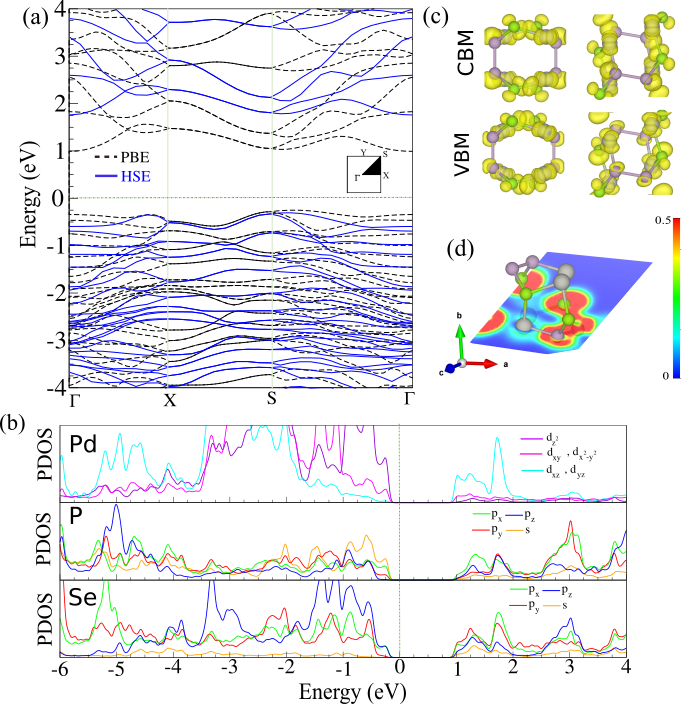}
	\caption{ (a) Electronic band structure and (b) PDOS and (c) electron localization function (ELF) of PdPSe monolayer. The zero of energy is set to Fermi-level.}
	\label{3}
\end{figure}

\begin{table*}
\centering
\caption{\label{table1} 
Structural, electronic and mechanical parameters of PdPSe monolayer, as shown in Fig. \ref{1}, including: lattice constants $\textit{a,b}$; 
bond lengths between P-Se ($d_{1}$), Pd-Se ($d_{2}$), Pd-P ($d_{3}$), and Se-Se ($d_{4}$), 
bond angles between Pd-Se-Pd ($\theta_{1}$), P-Se-P ($\theta_{2}$), P-Se-Pd ($\theta_{2}$) and P-Se-Se ($\theta_{3}$), 
buckling ($t$); 
cohesive energy per atom, $(E_{coh})$; 
charge transfer $\Delta{Q}$;
work function $\Phi$;
band gap $E_{g}$ with PBE (HSE06); 
bulk modulus (B); shear modulus (S); Young's modulus (Y); and Poisson's ratio ($\nu)$) respectively.
}
\begin{tabular}{lccccccccccccccc} 
\hline\hline
Sys.&$\textit{a,b}$   & ${d_{1,2,3,4}}$  & ${\theta}{_{1,2,3}}$ & $t$  &$E_{coh}$  & $\Delta{Q}$ & $\Phi$ & $E_{g}$& B & S & Y& $\nu$\\  
& (\AA)         & (\AA)        & ($^{\circ}$)     &  \AA & (eV/atom) & (e)         & (eV)   & (eV)   & GPa & GPa & GPa &    \\
\hline
PdPSe         & 5.86,5.79   & 2.85,2.48,2.31,2.20 & 114.86,103.83,112.19 & 1.46 &-0.18& 0.18 & 5.90 & 1.40 (2.07) & 21.43&15.36 & 37.19&0.21 \\
\hline\hline
\end{tabular}
\end{table*}
 
The atomic structure of the PdPSe monolayer in the diverse views is represented in Figs. \ref{1}(a).
In the PdPSe, the lattice with a pentagonal primitive unit cell is determined by eight atoms (2Pd, 2 P, and 2 Se), and the primitive unit cell is shown as a red square.
The pentagonal lattice of PdPSe belongs to the $Pbca$ (60) space group with a unit cell lattice constants of 5.86 \AA{} ($\textit{a}$) and 5.79 \AA{} ($\textit{b}$). 
The bond lengths are determined to be: 2.85 \AA{} ($d_{1}$), 2.48 \AA{} ($d_{2}$), 2.31\AA{} ($d_{3}$) and 2.20 \AA{} ($d_{4}$).
While the bond angles of $\theta_{2}$), $\theta_{2}$ and $\theta_{3}$ are calculated 114.86, 103.83 and 112.19 $^{\circ}$,respectively. 
The PdPSe monolayer is not a planar structure. The buckling is determined as 1.46 \AA{}. The structural parameters are summarized in Table I.

The cohesive energy per atom, the energy obtained by adjusting the atoms in a crystalline compared to the gas state, that is quantifying the stability of materials, was computed applying the following equation:
\begin{equation}
E_{coh} = \frac{E_{tot}-(n_{Pd}E_{Pd}+(n_{P}E_{P}+n_{Se}E_{Se})}{n_{Pd}+n_{P}+n_{Se}},
\end{equation}
where $E_{Pd}$, $E_{P}$ and $E_{Se}$ represent the energies of isolated single Pd, P and Se atoms; $E_{tot}$ shows the total energy of the PdPSe. 
Additionally, n$_{Pd}$, n$_{P}$, and n$_{Se}$ determines the number of Pd, P, and Se atoms in the primitive unit cell, respectively.
Our results show that the cohesive energies are -0.18 eV/atom.
The negative cohesive energy obtained for this layer suggests that the free-standing state of the nanostructure remains likely to be stable.

The difference charge density showed in Fig. \ref{1}(b) is specified as: 
\begin{equation}
\Delta\rho =\rho_{tot}-\rho_{Pd}-\rho_{P}-\rho_{Se},
\end{equation} 
where $\rho_{tot}$, $\rho_{Pd}$, $\rho_{P}$ and $\rho_{Se}$ depicts the charge densities of the isolated atoms and PdPSe monolayer, respectively. 
The charge depletion and accumulation are illustrated by coloring schemes with yellow and blue regions, correspondingly. 
C and N atoms are negatively charged and are enclosed by the positively charged B atoms.  
Based on the charge transfer ($\Delta \rho$) analysis, the Se atoms gain as 0.18 from the nearby Pd and P atoms.
The electrostatic potential of the PdPSe monolayer is flat in the vacuum area (Fig. \ref{1}(c)).
Applying the formula $\Phi =E_{vacuum}-E_{F}$, the work function of the PdPSe was defined as 5.90 eV. 

\subsection{Dynamics and thermal stability}
	
The phonon dispersion spectrum through the whole BZ is shown in Figs. \ref{2}(a). The phonon spectrum represents an in-plane transverse (TA) and a longitudinal (LA) optical mode, with a linear dispersion, and an out-of-plane flexure mode (ZA) with quadratic dispersion in the long-wavelength limit. The quadratic dispersion results in the large density of modes for ZA flexural phonons, which carry the most heat comparison LA and TA phonon modes. We may expect the other way around because the flexural acoustic ZA phonons have vanishing group velocities for a wave vector near zero.  The selection rule for three-phonon scattering can also explain this unexpected result, which strongly restricts the phase space for this scattering \cite{a1}. Flexural modes play a vital role in contributing to the thermal conductivity of graphene and structures with flexural modes, both as carriers \cite{a1} and as scatterers \cite{a2}. Flexural modes are perceived in the eminent 2DMs such as graphene\cite{a1}, hexagonal boron nitride (h-BN)\cite{a3}, and arsenene\cite{a4}.

The phonon dispersion spectrum is free from any imaginary frequency, discovering the dynamical stability of the PdPSe monolayer. In addition, a remarkable gap between the high-frequency and low-frequency optical branches is present. The splitting between the longitudinal and the transversal optical phonon frequencies (LO-TO splitting) presented at the $\Gamma$ point predicts the effects of the dipole-dipole interactions and the high-frequency dielectric constants \cite{a5}. 

We analyze the thermal stability of the PdPSe monolayer by the \textit{ab initio} molecular dynamics (AIMD) trajectories at 600 K. The AIMD computational results for the PdPSe monolayer at 600 K is illustrated in Fig. \ref{2}(b). 
The inset in the panel of Fig. \ref{2}(b) demonstrates the optimized structure after five ps of simulation. Analysis of the AIMD trajectories shows that the 2D structure could stay intact at 600 K, with very stable temperature and energy profiles, proving the thermal stability of the PdPSe monolayer which is beneficial for the experimental synthesis of monolayer PdPSe.

Next, we investigate the mechanical stability and properties of the PdPSe. The linear elastic constants are calculated using the harmonic approximation. We find that the elastic parameters agree with the twelve Born criteria \cite{mech1}. We calculate the bulk (B), shear (S) and Young’s moduli (Y) of PdPSe monolayer as 21.43 GPa, 15.36 GPa, and 37.19, respectively. The Poisson's ratio ($\nu)$=0.21) suggests that the PdPSe monolayer is a brittle structure as it is less than 0.33 \cite{A33}.

\subsection{Electronic properties}

\begin{figure}[!b]
\centering
\includegraphics[scale=1]{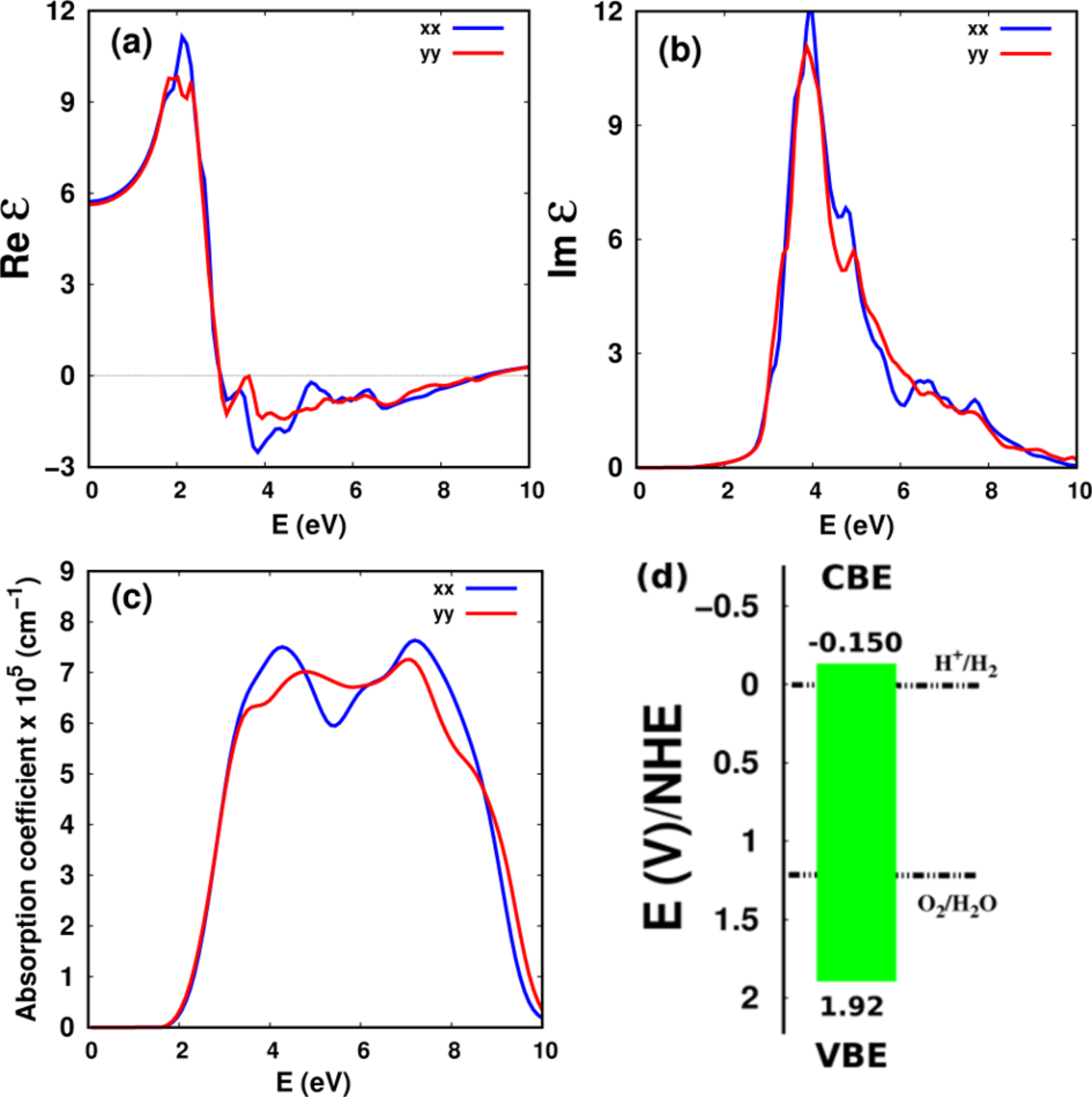}
\caption{(a) Real, (b) imaginary parts of the dielectric function, (c) the optical absorption spectrum as a function ofas a function of photon energy for PdPSe monolayer using the RPA+HSE06 approaches and (d) Band alignments of PdPSe monolayer for photocatalytic water splitting and carbon dioxide reduction. The band edges are given with respect to the NHE (normal hydrogen electrode) potential (in Volts).}
\label{4}
\end{figure}

The electronic band structure show that the PdPSe monolayer is an indirect bandgap semiconductor with a forbidden gap, E$_g$, of 1.26 eV and 2.07 eV within the PBE and HSE06 functional, respectively (see Fig.~\ref{3}).
The HSE06 hybrid functional does not alter the type of indirect bandgap in the PdPSe monolayer. 
Notably, the band gap of PdPSe is similar to that of MoS$_2$ and WS$_2$ monolayers \cite{tmd_bg}.
The valance band minimum (VBM) and conduction band maximum (CBM) are located at the $\Gamma$ point and S point, respectively. The VBM and the CBM are almost isotropic around the $\Gamma$ point. 
Based on the partial density of state (PDOS) analysis, as depicted in Fig.~\ref{3}(b), the VBM are composed mainly of the Pd-$d_{z^2}$-$d_{xy}$-$d_{x^2-y^2}$, P-$s, p_{x}$ and Se-$p_{x,z}$ orbitals states, while the CBM are originated from $d_{xz,yz}$, $p_{x,z}$ and $p_{x,y,z}$ of Pd, P and Se atoms of PdPSe, respectively. 

The electron localization function (ELF) along the (0 0 1) plane held the monolayer as depicted in Fig. \ref{3}(d). The red (blue) color means high (low) electron density in the ELF. ELF takes a value between 0 and 0.5, where ELF = 0.5 corresponds to the perfect localization. Electron localization occurs on the Se atoms. The maximum ELF value between Pd and P/Se atoms is approximately 0.50, representing a large proportion of the covalent bond in the PdPSe monolayer.

\begin{figure}
	\includegraphics[scale=1]{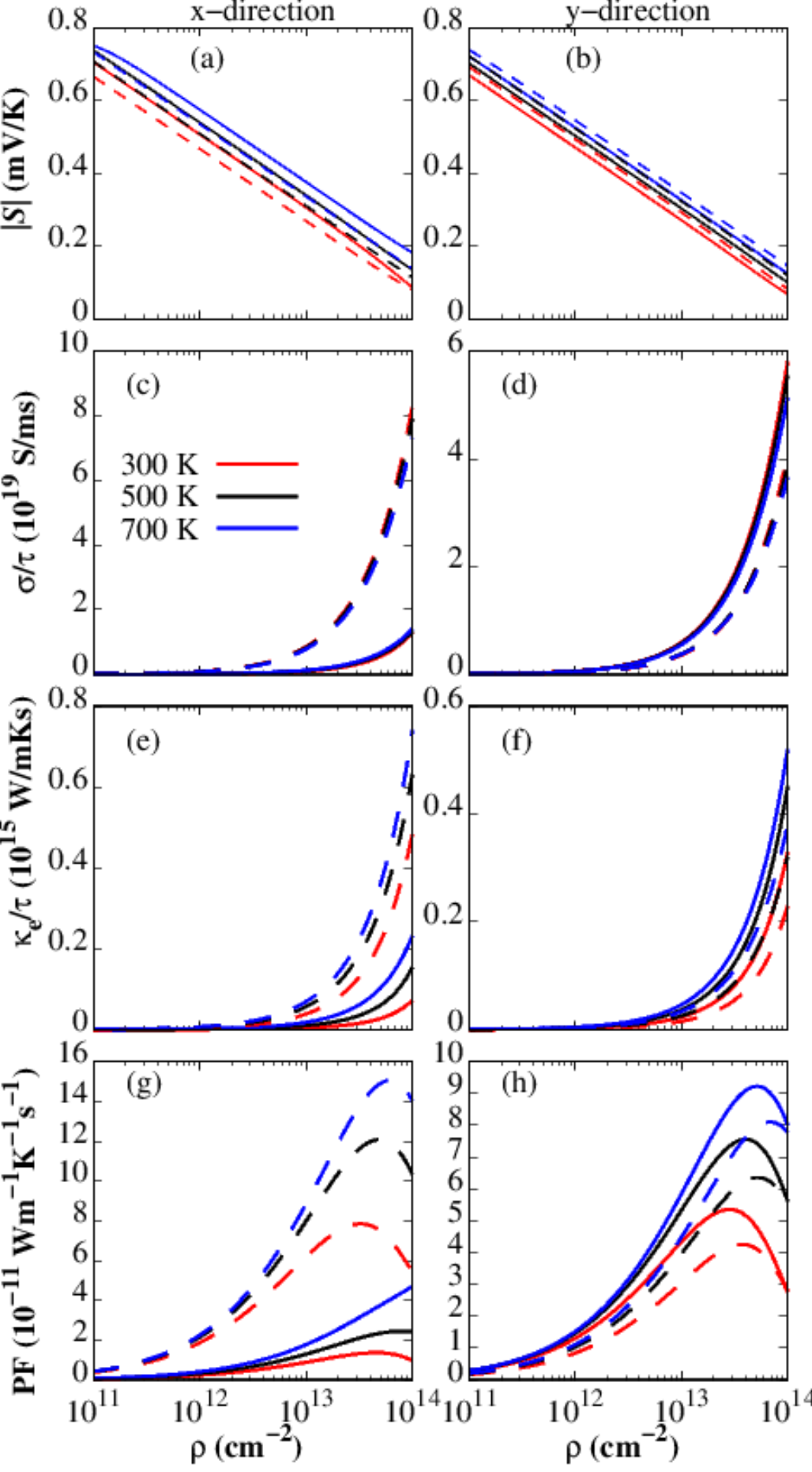}
	\caption{ The electronic transport coefficients of the PdPSe monolayer as function carrier concentration at 300 K, 500 K, and 700 K, (a,b) Seebeck coefficient ($S$), (c,d) relaxation time-scaled electrical conductivity ($\sigma/\tau$), (e,f) electronic thermal conductivity ($\kappa_{e}/\tau$), and (g,h) power factor ($PF$). The solid and dashed lines represent the $p$-type and $n$-type doping, respectively.}
	\label{5}
\end{figure}

\subsection{Optical properties and water splitting}

We now investigate the optical properties of the PdPSe monolayer. The dielectric function (the real and imaginary parts) is calculated using the RPA method over the HSE06. Figs. \ref{4}(a,b) shows the  real and imaginary parts of the dielectric function which describe the interaction between the electromagnetic field and monolayer PdPSe monolayer. 
The static value of the dielectric constant is the value of the real part at the zero energy which is found to be 5.73 and 5.63 for the xx and yy components, respectively. The dielectric function in the zz direction is very small as compared to the corresponding function in xx and yy components due to the monolayered nature of PdPSe. The dielectric function is anisotropic along the xx, yy, and zz directions. 
The first plasma frequencies appear at $\sim$ 3.01 eV and 2.97 eV along xx and yy direction, respectively. Other plasma frequencies appear at 8.95 eV and 9.10 eV along xx and yy direction, respectively.
The imaginary part of the dielectric function represents the sum of all transitions from the valence bands to the conduction bands. It starts to increase above $\sim$ 2.1 eV which is related to the band gap value. The imaginary part increases from 2.1 eV to 4 eV which are in the visible and the ultraviolet ranges of light. Figure \ref{4}(c) shows the absorption spectrum for PdPSe monolayer. The onset of significant absorption occurs at $\sim$ 2.0 eV which is related to the onset of the imaginary part and the band gap of PdPSe. The absorption spectrum suggests that the PdPSe has the ability to absorb the visible and ultraviolet light range.
Fig. \ref{4}(d) illustrates the band alignment of the oxidation and reduction potentials for water splitting with respect to the band edges of PdPSe monolayer suggesting that it is a potential candidate for photocatalytic water splitting. Most importantly, PdPSe monolayer exhibits a suitable band gap $\sim$ 2.07 eV and the band edges CBE and VBE must be higher (more negative) and lower (more positive) than the hydrogen reduction potential of H$_{+}$/H$_{2}$ and the water oxidation potential of H$_{2}$O/O$_{2}$, respectively, for efficient water splitting to take place. The CBE is computed from the relation $E_{CBE} = X-0.5 E_{gap}-4.5$ eV \cite{AAA1,AAA2,AAA3}, then the valence band edge is calculated from $E_{VBE} = E_{CBE}+E_{gap}$, where $X$ is the geometric mean of the electro-negativity of the ingredient atoms \cite{AAA4} and $4.5$ eV is the free energy of the electron (with respect to the vacuum level).

\subsection{Thermoelectric properties}

The calculated $S$, $\sigma$, $\kappa_{e}$ and power factor ($PF=S^{2}\sigma/\begin{tabular}{c} \ensuremath{\tau}\end{tabular}$) for both $p$- and $n$-type for monolayer PdPSe are shown in Fig. \ref{5}. The $S$ linearly decreases (increases) with increasing carrier concentration (temperature) and which is agreement with  previous work \cite{sofo_mahan}.
At 300 K, the $S$ for the hole (electron) doping of 10$^{11}$ are 702 (668) and 669 (692) $\mu$VK$^{-1}$ along the $x$- and $y$-direction, respectively. The $S$ does not varry significantly with both doping and transport directions. In general, $\sigma/\tau$ and $\kappa_{e}/\tau$ increase with an increasing carrier concentration. The $\sigma/\tau$ for the holes along the $x$-direction is lower than that of the electrons due to the flat valence band while that of both electron and hole are similar along the $y$-direction. The $PF$ is related to the interplay between $S$ and $\sigma/\tau$. As shown in Figs. 5(g) and (f), the $PF$ increases with an increasing carrier concentration initially to reach a peak before decreases with higher carrier concentration. The maximum $PF$ values of 7.9 and 5.2 are achieved for $n$-type along the $x$-direction and $p$-type along $y$-direction, respectively. Such high values of $PF$ demonstrating that PdPSe is
a promising material for thermoelectricity.

\section{Conclusion}

In summary, first-principle investigations is carried out study the physical properties of \textit{penta}-PdPSe monolayer. The stability is confirmed based on the dynamical, phonon dispersions and mechanical studies. Flexural modes and LO-TO splitting are observed in the phonon dispersion of the PdPSe monolayer. The monolayer has brittle behaviour. The electronic structure of the PdPSe monolayer calculated using the HSE06 reveals that the monolayer is an indirect semiconductor with a bandgap of 2.07 eV. The optical properties reveal the PdPSe monolayer can absorb the visible and ultraviolet the light. 
High $PF$ is achievable at room temperature.
Furthermore, \textit{penta}-PdPSe is predicted to be a photocatalytic material for water splitting application. Our results highlight the strong application potential of \textit{penta}-PdPSe monolayer. 

\section{Conflicts of interest}
The authors declare that there are no conflicts of interest regarding the publication of this paper.


\section{ACKNOWLEDGMENTS}
This work was supported by the National Research Foundation of Korea (NRF) grant funded by the Korea government (MSIT) (NRF-2015M2B2A4033123). 



\begin{thebibliography}{}

\end{thebibliography}


\begin{thebibliography}{100}

\bibitem{1}	
A. Bafekry, S.F. Shayesteh, F.M. Peeters, 
J. Appl. Phys. 126 (2019). 

\bibitem{2}	
A. Bafekry, C. Stampfl, M. Faraji, M. Yagmurcukardes, M.M. Fadlallah, H.R. Jappor, M. Ghergherehchi, S.A.H. Feghhi, 
Appl. Phys. Lett. 118 (2021) 203103. 

\bibitem{3}	
A. Bafekry, I. Abdolhosseini Sarsari, M. Faraji, M.M. Fadlallah, H.R. Jappor, S. Karbasizadeh, V. Nguyen, M. Ghergherehchi, 
Appl. Phys. Lett. 118 (2021) 143102.

\bibitem{4}	
A.O. Muhsen Almayyali, H.O. Muhsen, M. Merdan, M.M. Obeid, H.R. Jappor, 
Two-dimensional ZnI2 monolayer as a photocatalyst for water splitting and improvement its electronic and optical properties by strains, 
Phys. E Low-Dimensional Syst. Nanostructures. 126 (2021) 114487. 

\bibitem{5}	
A. Bafekry, B. Mortazavi, M. Faraji, M. Shahrokhi, A. Shafique, H.R. Jappor, C. Nguyen, M. Ghergherehchi, S.A.H. Feghhi, 
Sci. Rep. 11 (2021) 10366. 

\bibitem{6}	
S. Qin, Q. Du, R. Dong, X. Yan, Y. Liu, W. Wang, F. Wang, Robust, 
Carbon N. Y. 167 (2020) 668-674.

\bibitem{7}	
S.C. Liufu, L.D. Chen, Q. Yao, C.F. Wang, 
Appl. Phys. Lett. 90 (2007) 112106.

\bibitem{8}	
M.M. Obeid, A. Bafekry, S. Ur Rehman, C. V. Nguyen, 
Appl. Surf. Sci. 534 (2020) 147607.

\bibitem{9}	
K.S. Novoselov, A.K. Geim, S. V. Morozov, D.A. Jiang, Y.Y. Zhang, S. V. Dubonos, A.A. Firsov, I. V. Grigorieva, A.A. Firsov, 
Sci. 306 (2004) 666-669. 

\bibitem{10}	
J.D. Wood, S.A. Wells, D. Jariwala, K.S. Chen, E. Cho, V.K. Sangwan, X. Liu, L.J. Lauhon, T.J. Marks, M.C. Hersam, 
Nano Lett. 14 (2014) 6964-6970. 

\bibitem{11}	
B. Lalmi, H. Oughaddou, H. Enriquez, A. Kara, S. Vizzini, B. Ealet, B. Aufray, 
Appl. Phys. Lett. 97 (2010) 223109. 

\bibitem{12}	
J. Ji, X. Song, J. Liu, Z. Yan, C. Huo, S. Zhang, M. Su, L. Liao, W. Wang, Z. Ni, Y. Hao, H. Zeng, 
Nat. Commun. 7 (2016) 13352.

\bibitem{13}	
A.O.M. Almayyali, B.B. Kadhim, H.R. Jappor, 
Chem. Phys. 532 (2020) 110679. 

\bibitem{14}	
R. Chaurasiya, A. Dixit, 
Phys. Chem. Chem. Phys. 22 (2020) 13903-13922.

\bibitem{15}	
Y. Su, M.A. Ebrish, E.J. Olson, S.J. Koester, 
Appl. Phys. Lett. 103 (2013) 263104. 

\bibitem{16}	
B.Z. Sun, Z. Ma, C. He, K. Wu,
Phys. Chem. Chem. Phys. 17 (2015) 29844-29853.

\bibitem{17}	
Y. Huang, E. Sutter, J.T. Sadowski, M. Cotlet, O.L.A. Monti, D.A. Racke, M.R. Neupane, D. Wickramaratne, R.K. Lake, B.A. Parkinson, P. Sutter, 
ACS Nano. 8 (2014) 10743-10755. 


\bibitem{18}	
A. Bafekry, C. Stampfl, B. Akgenc, B. Mortazavi, M. Ghergherehchi, C. V. Nguyen, 
Phys. Chem. Chem. Phys. 22 (2020) 6418-6433.

\bibitem{19}	
A. Bafekry, M. Shahrokhi, H.R. Jappor, A. Shafique, F. Shojaei, S.A.H. Feghhi, M. Ghergherehchi, D. Gogova, 
Nanotechnol. 32 (2020) 215702. 


\bibitem{20}	
A. Bafekry, M. Yagmurcukardes, B. Akgenc, M. Ghergherehchi, C. V. Nguyen, 
J. Phys. D. Appl. Phys. 53 (2020) 355106. 

\bibitem{21}	
H.D. Bui, H.R. Jappor, N.N. Hieu, 

\bibitem{22}	
I. Ronneberger, Z. Zanolli, M. Wuttig, R. Mazzarello, 
Adv. Mater. 32 (2020) 2001033.

\bibitem{23}	
X. Tan, X. Tan, G. Liu, J. Xu, H. Shao, H. Hu, M. Jin, H. Jiang, J. Jiang, 
J. Mater. Chem. C. 5 (2017) 7504-7509.

\bibitem{24}	
H.R. Jappor, M.A. Habeeb, 
Curr. Appl. Phys. 18 (2018) 673-680. 

\bibitem{aa1} 
Y.-L. Hong, Zhibo Liu, L. Wang, T. Zhou, W. Ma, C. Xu, S. Feng, L. Chen, M.-L. Chen ,D.-M. Sun, X.-Q. Chen, H.-M. Cheng, W. Ren, 
Sci. 369 (2020) 670-674.

\bibitem{aa2} 
A. Bafekry, M. Faraji, D.M. Hoat, M.M. Fadlallah, M. Shahrokhi, F. Shojaei, D. Gogova, M. Ghergherehchi, 
J. Phys. D: Appl. Phys. 54 (2021) 155303. 

\bibitem{aa3} 
Q. Wang, L. Cao, Sh.-J. Liang, W. Wu, G. Wang, Ch. H. Lee, W. L. Ong, H. Y. Yang, L. K. Ang, Sh. A. Yang,  Y. Sin Ang,
npj 2D Mater. Appl. 5 (2021) 71.

\bibitem{aa4} 
L. Cao, G. Zhou, Q. Wang, L. K. Ang, Y. S. Ang, 
Appl. Phys. Lett. 118 (2021), 013106 (2021).


\bibitem{25}	
T.A. Bither, P.C. Donohue, H.S. Young, 
J. Solid State Chem. 3 (1971) 300-307. 

\bibitem{26}	
W. Jeitschko, 
Acta Crystallogr. Sect. B. 30 (1974) 2565-2572. 

\bibitem{27}	
J.K. Burdett and B. A. Coddens, 
Inorg. Chem. 27 (2002) 418-421.

\bibitem{28}	
A. Hamidani and B. Bennecer, 
Comput. Mater. Sci. 48 (2010) 115-123. https://doi.org/10.1016/J.COMMATSCI.2009.12.017.

\bibitem{29}	
J. V. Marzik, R. Kershaw, K. Dwight, A. Wold, 
J. Solid State Chem. 44 (1982) 382-387. 

\bibitem{30}	
J.C.W. Folmer, J.A. Turner, B.A. Parkinson, 
J. Solid State Chem. 68 (1987) 28-37.

\bibitem{31}	
Y. Jing, Y. Ma, Y. Wang, Y. Li, T. Heine, 
Chem. A Eur. J. 23 (2017) 13612-13616.

\bibitem{32}	
Y. Ma, L. Kou, X. Li, Y. Dai, T. Heine, 
NPG Asia Mater. 2016 84. 8 (2016) e264-e264.

\bibitem{33}	
S. Zhang, J. Zhou, Q. Wang, X. Chen, Y. Kawazoe, P. Jena, 
Proc. Natl. Acad. Sci. U. S. A. 112 (2015) 2372-2377. 

\bibitem{34}	
M. Naseri, 
Appl. Surf. Sci. 423 (2017) 566-570. 

\bibitem{35}	
M. Yagmurcukardes, H. Sahin, J. Kang, E. Torun, F.M. Peeters, R.T. Senger, 
J. Appl. Phys. 118 (2015) 104303. 

\bibitem{36}	
Z. Azarmi, M. Naseri, S. Parsamehr, 
Chem. Phys. Lett. 728 (2019) 136-141. 

\bibitem{37}	
A.D. Oyedele, S. Yang, L. Liang, A.A. Puretzky, K. Wang, J. Zhang, P. Yu, P.R. Pudasaini, A.W. Ghosh, Z. Liu, C.M. Rouleau, B.G. Sumpter, M.F. Chisholm, W. Zhou, P.D. Rack, D.B. Geohegan, K. Xiao, 
J. Am. Chem. Soc. 139 (2017) 14090-14097. 

\bibitem{38}	
P. Li, J. Zhang, C. Zhu, W. Shen, C. Hu, W. Fu, L. Yan, L. Zhou, L. Zheng, H. Lei, Z. Liu, W. Zhao, P. Gao, P. Yu, G. Yang, 
Adv. Mater. (2021) 2102541. 

\bibitem{GGA-PBE1} 
J. P. Perdew, K. Burke, and M. Ernzerhof, 
Phys. Rev. Lett. 77, 3865 (1996).


\bibitem{GGA-PBE2} 
J. P. Perdew, K. Burke, and M. Ernzerhof,
Phys. Rev. Lett. 78, 1396 (1997).

\bibitem{vasp1} 
G. Kresse and J. Hafner, 
Phys. Rev. B 47, 558 (1993).

\bibitem{vasp2} 
G. Kresse and J. Hafner,
Phys. Rev. B 49, 14251 (1994).

\bibitem{hse} 
J. Heyd, G. E. Scuseria, and M. Ernzerhof,
J. Chem. Phys. 118, 8207 (2003).


\bibitem{Monkhorst} 
H.J. Monkhorst and J.D. Pack,
Phys. Rev. B 13, 12, (1976).

\bibitem{Henkelman} 
G. Henkelman, A. Arnaldsson, and H. Jonsson,
Comput. Mater. Sci. 36, 354 (2006).


\bibitem{Grimme}  
S. J. Grimme, 
Comput. Chem. 27, 1787 (2006).

\bibitem{phon} 
D. Alfe, 
Comput. Phys. Commun. 180, 2622 (2009).

\bibitem{a1}
L. Lindsay, D. A. Broido, and N. Mingo, 
Phys. Rev. B 82, 115427 (2010).

\bibitem{a2}
L. F. C. Pereira and D. Donadio, 
Phys. Rev. B 87, 125424 (2013).


\bibitem{a3}
[41] L. Lindsay and D. A. Broido, 
Phys. Rev. B 84, 155421 (2011).

\bibitem{a4}
Majid Zeraati, S. Mehdi Vaez Allaei, I. Abdolhosseini Sarsari, Mahdi Pourfath, and Davide Donadio,
Phys. Rev. B 93, 085424 (2016).

\bibitem{a5}
Kunpeng Yuan, Zhehao sun, Xiaoliang Zhang, and Dawei tang
Scientific Reports 9 (2019) 9490


\bibitem{DFT-D3}
T. Bucko, J. Hafner, S. Lebegue and J. G. Angyan,
Phys. Chem. A 114 (2010) 118145.


\bibitem{Tersoff}
J. Tersoff and D. R. Hamann. 
Phys. Rev. Lett. 50 (1983) 1998-2001.

\bibitem{WSxM}
I. Horcas, R. Fernndez, J. M. Gmez-Rodrguez, J. Colchero, J. Gmez-Herrero, and A. M. Baro. 
Rev. Scient. Inst. 78 (2007) 013705.

\bibitem{boltztrap2} G. K. H. Madsen, J. Carrete, and M. J. Verstraete,
Comput. Phys. Commun. 231 {2018} 140-145.

\bibitem{mech1}
F. Mouhat and F.-X. Coudert,
Phys. Rev. B 90, (2014) 224104. 

\bibitem{A33} W. C. Hu, Y. Liu, D. J. Li, X. Q. Zeng and C. S. Xu, 
Comput. Mater. Sci. 83, (2014) 27.

\bibitem{tmd_bd} J. Gusakova et al, Phys. Stat. Solidi a 214 (2017), 1700218. 

\bibitem{AAA1}
H. Yu, S. Ouyang, S. Yan, Z. Li, T. Yu, and Z. Zou,
J. Mater. Chem., 21 (2011) 11347.


\bibitem{AAA2}
M. M. Fadlallah and U. Eckern,
physica status solidi (b) 257, (2020) 1900217. 

\bibitem{AAA3}
A. A. Maarouf, D. Gogova and M. M. Fadlallah,
Applied Physics Letters, 119 (2021) 063901.

\bibitem{AAA4}
L. J. Bartolotti, S. R. Gadre, and R. G. Parr,
J. ACS, 102 (1980) 2945.

\bibitem{A10}
F. Wooten, 
Academic Press, 2013.

\bibitem{sofo_mahan} 
J. O. Sofo, G. D. Mahan, 
Phys. Rev. B 58 (1998), 15620. 

\end{thebibliography}
\end{document}